\documentclass{article}

\usepackage[preprint]{neurips_2024}

\usepackage[utf8]{inputenc} 
\usepackage[T1]{fontenc}    
\usepackage{hyperref}       
\usepackage{url}            
\usepackage{booktabs}       
\usepackage{amsfonts}       
\usepackage{graphicx}       
\usepackage{nicefrac}       
\usepackage{microtype}      
\usepackage[dvipsnames]{xcolor}         
\usepackage{xspace}
\usepackage{float}
\usepackage{natbib}

\title{Devstral: Fine-tuning Language Models\\ for Coding Agent Applications}

\newcommand{\ours}{\texttt{Devstral-Small}\xspace}
\newcommand{\oursSmall}{\texttt{Devstral-Small}\xspace}
\newcommand{\oursJuly}{\texttt{Devstral-Small-2507}}

\newcommand{\openhands}{\texttt{OpenHands}\xspace}

\newcommand{\swebv}{\texttt{SWE-bench Verified}\xspace}

\newcommand{\sref}[1]{\S\ref{#1}}

\usepackage{hyperref}
\definecolor{scholarblue}{rgb}{0.21,0.49,0.74}
\hypersetup{
	colorlinks=true,
	linkcolor=red,
	filecolor=blue,      
	urlcolor=scholarblue,
	citecolor=scholarblue,
}

\author{%
  Mistral AI \; $\mathcal{\times}$ \; All Hands AI
}
\usepackage{color-edits}

\begin{document}

\maketitle

\begin{abstract}
  We introduce \ours, a lightweight open source model for code agents with the best performance among models below 100B size. In this technical report, we give an overview of how we design and develop a model and craft specializations in \emph{agentic} software development.
  The resulting model, \oursSmall is a small 24B model, fast and easy to serve. Despite its size, \oursSmall still attains competitive performance compared to models more than an order of magnitude larger.
\end{abstract}

\section{Introduction}
Large language models~(LLMs) have been increasingly adapted for code-related tasks, demonstrating remarkable capabilities in generating, understanding, and completing code~\citep{austin2021programsynthesislargelanguage,chen2021evaluatinglargelanguagemodels,openai-gpt41,anthropic-claude4}.
Several high-performance open code completion models\textemdash such as CodeLlama~\citep{roziere2023code}, Qwen 2.5 Coder~\citep{hui2024qwen2} or Codestral~\citep{codestral2501}\textemdash were released for code completion and generation tasks. 

In the past year, the coding agent has emerged as a unique and novel application of LLM code LLMs~\citep{devin,yang2024sweagent,wang2025openhandsopenplatformai}:
the integration of LLMs into agent-based systems has opened new avenues for automating complex software engineering workflows, such as solving issues, performing refactors or implementing features involving several files in a codebase. Agentic coding is, however, challenging; previous generations of models excel in generating syntactically correct code but often struggle with complex, multi-step programming tasks that require reasoning over project-specific contexts or external tools. 
Although closed models have made significant progress~\cite{anthropic-claude4,openai-gpt41}, open models still lack the agentic capabilities needed for iterative development processes, such as debugging or integrating with software tools.
Hence, code agents use mainly closed models such as Claude~\citep{anthropic-claude4}.

Addressing this gap, we introduce \oursSmall, a 24B model specialized for code agent applications.  By incorporating agentic reasoning, \ours possess the capabilities to interact with development environments and handle multi-step tasks more effectively. 
Despite having only 24B parameters, \oursSmall is a high-performing model for agentic tasks for code. 


\section{\oursSmall}

\subsection{Base Model}
\oursSmall is a dense Transformer consisting of a total of 24 billion parameters. It is based on Mistral Small 3, has 40 layers and uses grouped query attention.
%
The model is pre-trained on diverse sources of text including both natural language and code. A long context extension phase boosts the model context size to 128k tokens suitable for code agent and other long context tasks.

\subsection{Data}
\label{subsec:data}
For code agents, we want to foster an interaction pattern in which the agent, and consequently the underlying model, alternates between chain-of-thought \citep{wei2022chain} reasoning and actions within the coding environment to perform the software engineering task.
%
We generate supervised trajectories by running an agent in the SWE-Gym environments~\citep{pan2024trainingsoftwareengineeringagents} with the \openhands CodeAct scaffold \citep{wang2025openhandsopenplatformai,DBLP:conf/icml/WangCY0L0J24}. 
We execute unit tests to determine the final patch quality.

We also include a selected mixture of natural language data to retain general natural language understanding capabilities in \ours. 
%


\subsection{Post-training}
\label{subsec:post}

We employ a two-stage training process. Initially, we train the model on a larger subset of rollouts that meet our first level of quality standards, determined by a heuristic filter. In the second stage, we fine-tune the model using only the trajectories that pass our strictest filters.

We then perform additional rounds of rollouts with the finetuned model, and train further the model with policy optimization. 

\section{Experiments}
\label{sec:exp}
We show that \oursSmall achieves the state-of-the-art result among open models (\sref{sec:devstral-sota-open-source}). Furthermore, we conduct further analysis to examine the impact of the iterative evaluation protocol, maximum iteration limit, and the temperature used for sampling LLMs (see \sref{sec:model-analysis}).

\subsection{Setup}
\label{sec:exp-setup}
We evaluate \oursSmall using an agentic evaluation setup, where the model has access to a bash execution and a file edition tool. 
The agent explores and edits the code in a code base, mimicking the actions of a human software engineer. No retrieval tool or multi-sample voting is utilized.

\paragraph{Scaffold and Settings} 
We evaluate \ours using the \openhands scaffold \citep{wang2025openhandsopenplatformai}. 
\openhands is an open platform designed to develop AI agents that interact with the world similarly to human software developers. It provides a comprehensive framework that enables agents to write code, interact with command lines, and browse the web within a secure, sandboxed environment.
The platform features an event stream architecture that captures all agent actions and observations, allowing for detailed analysis and improvement of agent behavior.
%
This makes it an ideal scaffold for evaluating and comparing coding agent models such as \ours in a standardized environment.
All evaluation experiments in this paper are conducted using \openhands\footnote{at commit \texttt{dc8fc45e94e32498b026a6b6ea91aa0dcb2aa689} for reproducibility}. 
We utilize the built-in function calling template of \openhands \citep{openhands_fncallconverter} and disable the in-context learning examples for function calling.
We disable the web browsing feature of OpenHands during evaluation.

\paragraph{Benchmarks} We perform our main results and ablations on \swebv~\citep{swebenchverified}.
We compare \oursSmall to prominent open-source models available at the time of the release using the OpenHands scaffold~\citep{openhands2025benchmarksheet}.

\paragraph{Generation} For reproducibility, the results reported in this paper are obtained using the open-source vLLM~\citep{kwon2023efficient} framework. We use the online server optimized with paged attention and automatic prefix caching~\citep{kwon2023efficient}.

\paragraph{Iterative Evaluation Protocol}

We adopt the iterative evaluation protocol from \openhands \citep{openhands2024iterativeeval} to address the inherent stochasticity of agents and produce more consistent results.
Notably, even with temperature 0, sources of nondeterminism persist. The LLM backend may still exhibit slight stochasticity due to underlying hardware-level implementation details (e.g., GPU parallelism or floating-point rounding differences). In addition, agent runtime effects\textemdash such as differences in file system state or metadata like mtime (modification time)\textemdash can lead to variations in the input prompt depending on when the evaluation is executed. These subtle shifts can result in different outputs despite ostensibly identical initial conditions.

Instead of repeatedly running the same instance many times and averaging pass@1 estimates, our protocol achieves comparable robustness with far fewer executions. 
For each SWE-bench instance, we allow up to 3 independent attempts to generate a patch. Each attempt proceeds until the agent either completes the task with an \texttt{AgentFinishAction} or hits the iteration limit. The first attempt uses temperature=0, we raise it to 0.1 in the second and third attempts to introduce controlled variation and improve coverage.
By aggregating up to three varied attempts (with preference to later solutions), we obtain stable and reliable performance metrics at a fraction of the computational cost.
This makes the evaluation process significantly more efficient and scalable. We leave exploration of more advanced refinement strategies (e.g. \citet{madaan2023self_refine}) to future work.

\subsection{Results}
\label{sec:devstral-sota-open-source}

\begin{figure}[H]
    \centering
    \includegraphics[width=0.95\textwidth]{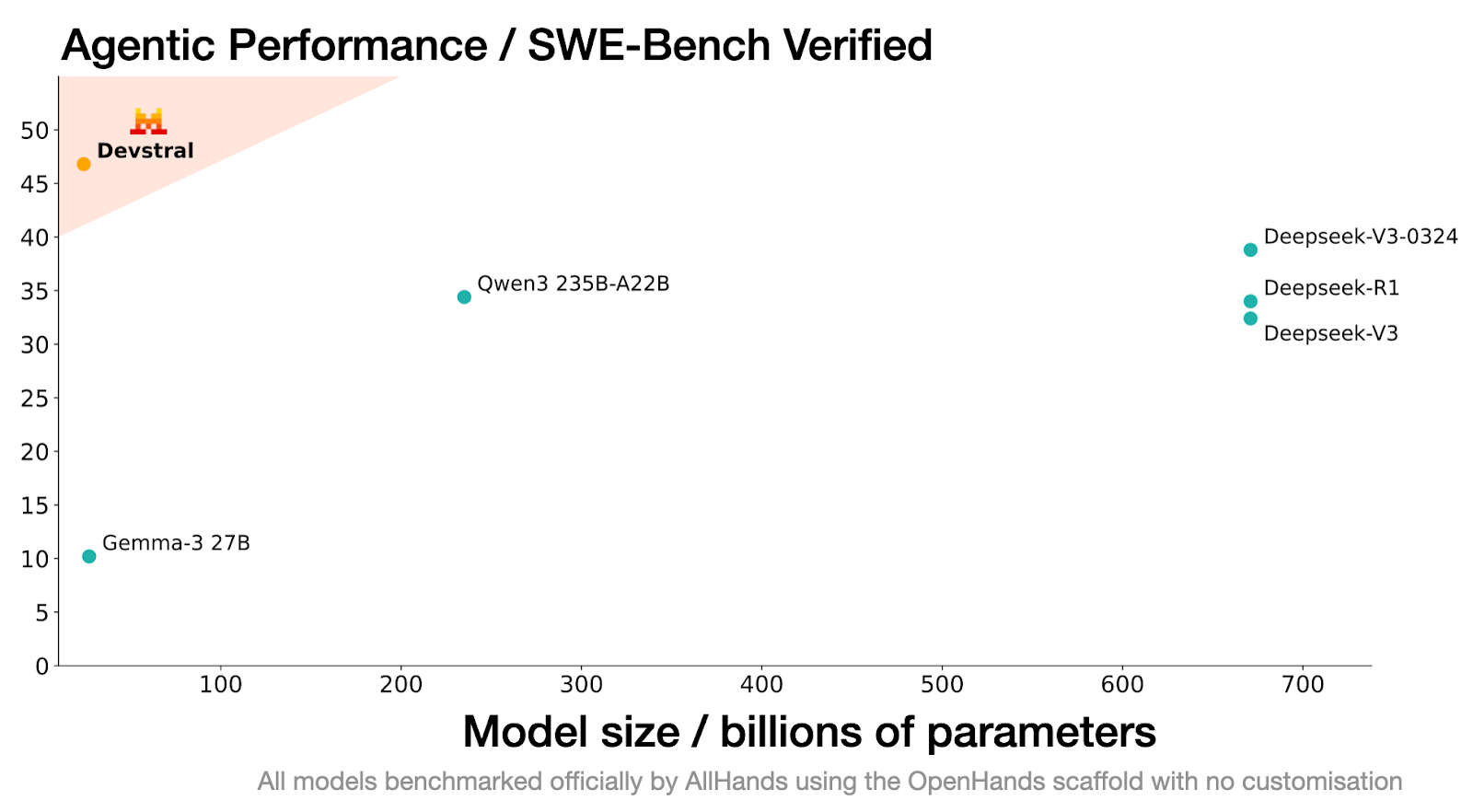}
    \caption{\oursSmall achieves state-of-the-art results among open models with the OpenHands scaffold. It outperforms models such as Qwen 3 235B and DeepSeek-V3 that are approximately 10 and 28 times larger respectively.}
    \label{fig:performance_oss}
\end{figure}

In Figure \autoref{fig:performance_oss}, we plot the performance of \oursSmall and other open models available at the time of the release against model size using the OpenHands scaffold and the same iterative evaluation protocol. \oursSmall outperforms other open models such as Qwen 3 and Deepseek V3 despite being a fraction of their size. 

These results highlight the effectiveness of specialized code agent models. Fundamentally, software engineering tasks differ from those in competitive programming, which has been the traditional focus of code generation models. By focusing on software engineering, \oursSmall achieves strong performance with only 24 billion parameters.

\begin{figure}[H]
    \centering
    \includegraphics[width=0.95\textwidth]{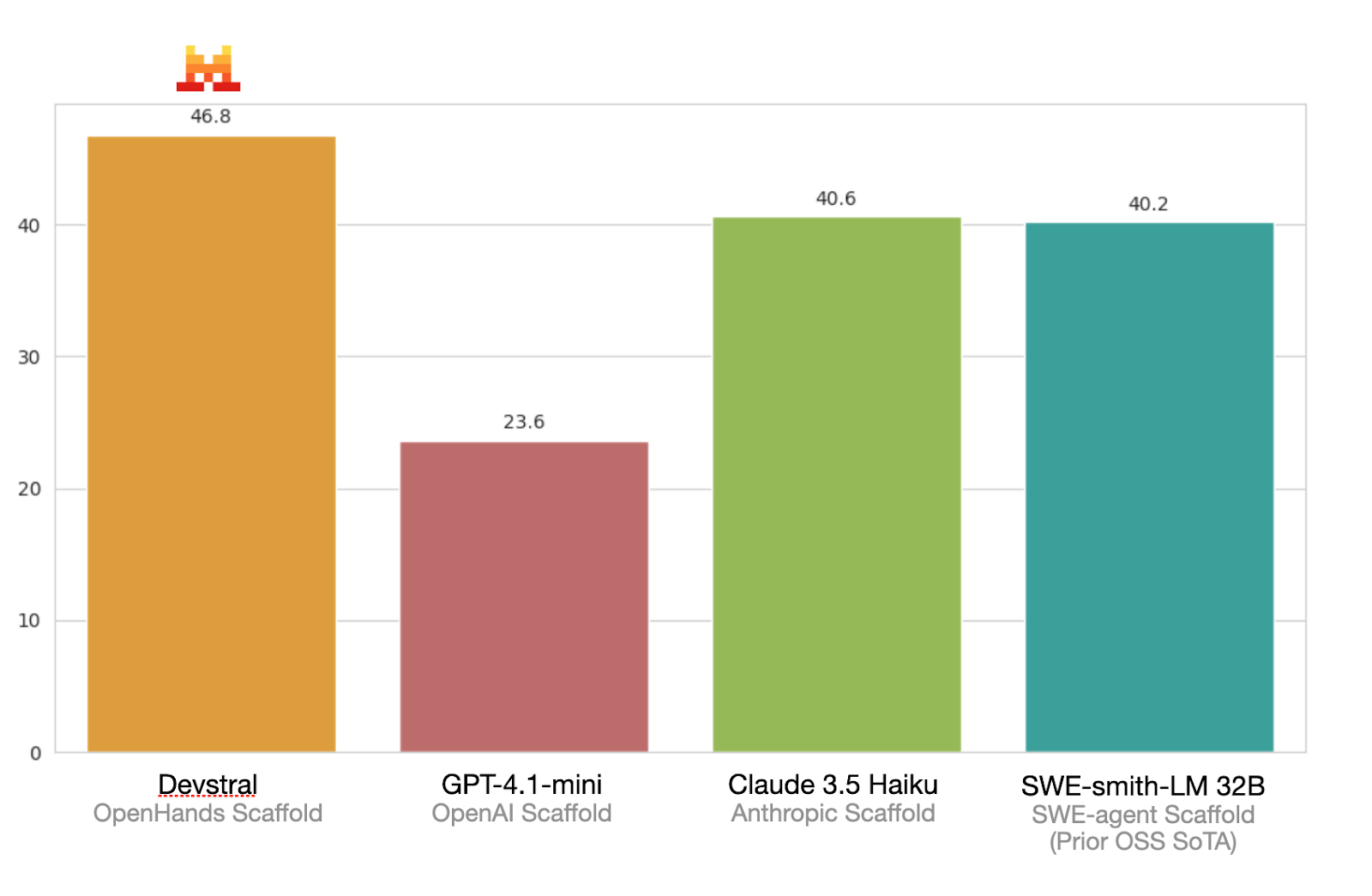}
    \caption{\textbf{\ours compared to other model on any scaffold.} We compare the performance of \ours to the reported SWE-Bench performance of GPT-4.1 mini and Claude 3.5 Haiku on custom scaffolds and to SWE-smith on the SWE-Agent scaffold.}
    \label{fig:performance_vs_small}
\end{figure}

In Figure \ref{fig:performance_vs_small}, we compare the performance of \oursSmall against both closed and open models, evaluated under various scaffolds, including those customized for their models. 
\oursSmall significantly outperforms the small models from OpenAI and Anthropic. For instance, \oursSmall exceeds the performance of the recent GPT-4.1-mini by over 20\%.

\subsection{Analysis}
\label{sec:model-analysis}
In this section, we conduct some extra experiments to provide more insight into the behavior of \ours. Specifically, we examine how the model behaves with different rollout horizons. Additionally, we present pass@N metrics at various temperatures to provide a view of the model's action landscape.

\subsubsection{Maximum Iteration Limits}
To understand how computational budget affects agent performance, we examined various maximum iteration limits during evaluation. We assessed our model under three different constraints: 30, 50, and 100 turns. These constraints represent different levels of computational resources allocated to each attempt at solving a github issue.

All these experiments are conducted with temperature 0 and employ the iterative evaluation protocol described in \autoref{sec:exp-setup}. This setup allows us to isolate the effect of the iteration limits on model performance while maintaining consistency across experimental conditions.

\autoref{tab:max_iterations} presents the results of our maximum iteration limit experiment. The results reveal several important insights about the relationship between computational budget and agent performance. Performance increases substantially from 30 to 50 iterations, with the resolve rate improving from 36.8\% to 46.8\%. However, further increasing the limit from 50 to 100 iterations yields no additional performance gains, with both configurations achieving identical 46.8\% resolve rates.

\begin{table}[H]
\centering
\begin{tabular}{cc}
\toprule
\textbf{Max Iterations} & \textbf{Resolve Rate (\%)} \\
\midrule
30 & 36.80 \\
50 & 46.80 \\
100 & 46.80 \\
\bottomrule
\end{tabular}
\caption{Performance of \ours under different maximum iteration limits. All experiments conducted with temperature=0 using the iterative evaluation protocol on \swebv.}
\label{tab:max_iterations}
\end{table}

These findings suggest that 50 iterations represent an optimal balance between computational efficiency and performance for \ours on software engineering tasks. The plateau in performance beyond 50 iterations indicates that the model typically either solves problems within this limit or encounters fundamental challenges that additional iterations cannot overcome.

\subsection{Temperature Scaling}
To investigate the impact of the sampling temperature on model performance, we conduct a comprehensive temperature scaling experiment using the \openhands scaffold. We evaluate \ours across four different temperature settings: $T = 0.1, 0.4, 0.7, 1.0$, representing a range from near-deterministic to highly stochastic generation. All experiments use a maximum iteration limit of 100 and do not employ the iterative evaluation protocol to isolate the effects of temperature variation.

We measure performance using the Pass@K metric, which evaluates whether the model succeeds in at least one of K attempts. This metric is particularly relevant for understanding how temperature affects the model's ability to generate successful solutions when given multiple opportunities.

\autoref{tab:temperature_scaling} presents the detailed Pass@K results across different temperature settings. The results reveal interesting patterns in how temperature affects model performance at different K values. Lower temperatures (T=0.1, T=0.4) tend to perform better at higher K values, suggesting that more deterministic generation benefits from multiple attempts. In contrast, higher temperatures show more variable performance, with T=1.0 achieving competitive results at Pass@4 despite lower initial performance.

\begin{table}[H]
\centering
\begin{tabular}{lcccc}
\toprule
\textbf{Temperature} & \textbf{Pass@1} & \textbf{Pass@2} & \textbf{Pass@3} & \textbf{Pass@4} \\
\midrule
T=0.1 & 43.2\% & 54.6\% & 58.8\% & 62.0\% \\
T=0.4 & 42.4\% & 52.2\% & 59.6\% & 62.2\% \\
T=0.7 & 43.2\% & 52.2\% & 55.4\% & 59.4\% \\
T=1.0 & 43.6\% & 52.0\% & 57.6\% & 60.4\% \\
\bottomrule
\end{tabular}
\caption{Temperature scaling experiment results showing Pass@K performance across different temperature settings. All experiments conducted with a maximum of 100 iterations without iterative evaluation protocol.}
\label{tab:temperature_scaling}
\end{table}

\begin{figure}[H]
    \centering
    \includegraphics[width=0.7\textwidth]{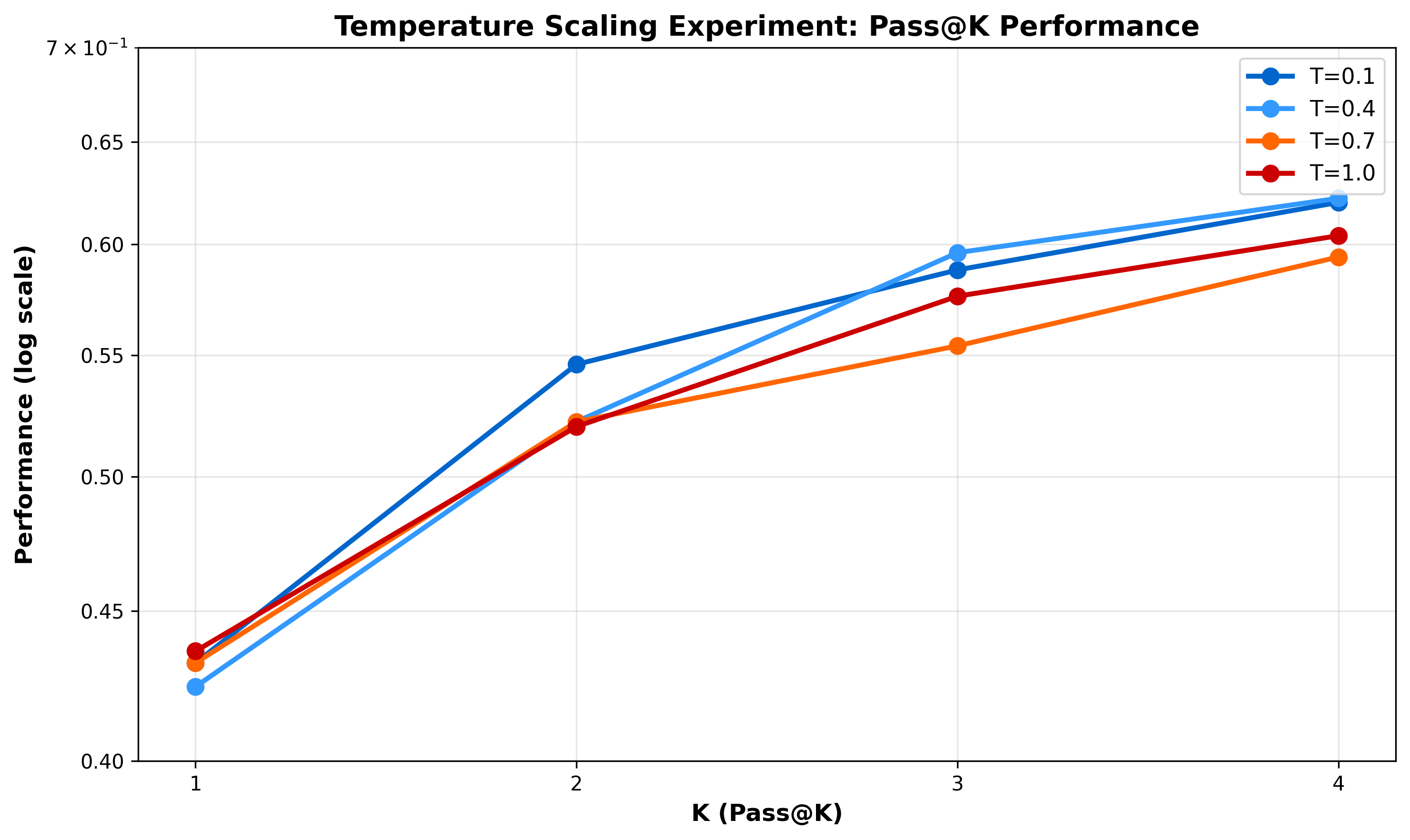}
    \caption{Temperature scaling experiment visualization showing Pass@K performance trends on a logarithmic scale. Lower temperatures (blue) consistently improve with increased K, while higher temperatures (red/orange) exhibit more variable performance patterns.}
    \label{fig:temperature_scaling}
\end{figure}

12:13amFigure \ref{fig:temperature_scaling} illustrates these results, clearly depicting performance trends for various values of  K and temperature settings. The figure shows that while all temperature settings achieve similar Pass@1 performance, around 43\%, the benefits of multiple attempts (higher K) vary significantly with temperature. Lower temperatures consistently improve with increased K, whereas higher temperatures display more volatile performance patterns.
These results are counter-intuitive, as higher temperatures are expected to scale better with K, as is the case for competitive programming exercises.

\subsubsection{Iterative Evaluation Protocol Ablation}
To validate the effectiveness of our iterative evaluation protocol, we present ablation results across the three iterations in \autoref{tab:iterative_ablation}. The results demonstrate the effectiveness of the iterative approach, showing consistent improvements in resolve rates across iterations while significantly reducing empty patch instances after the first iteration.

\begin{table}[H]
\centering
\begin{tabular}{cccc}
\toprule
\textbf{Iteration} & \textbf{Instances Run} & \textbf{Resolution Rate (\%)} & \textbf{Empty Patch Rate (\%)} \\
\midrule
1 & 500 & 43.0 & 7.6 \\
2 & 191 & 45.8 & 0.0 \\
3 & 121 & 46.8 & 0.0 \\
\bottomrule
\end{tabular}
\caption{Ablation results for the iterative evaluation protocol. The resolution rate is calculated as resolved instances divided by total instances (500). The empty patch rate represents the percentage of instances that resulted in empty patches. Instances run shows the number of instances processed in each iteration.}
\label{tab:iterative_ablation}
\end{table}


\section{Refreshing \ours Data}

After releasing the first generation of \ours, we developed an updated version by improving our dataset. 
We meticulously refined the data generation and curation process through careful ablation studies. To improve generalization to diverse scaffolds, we created a variety of pseudo-scaffolds. Our training included prompts in both XML and our native function calling formats. Furthermore, we fine-tuned the data filtering mechanism to ensure high-quality trajectory selection, which helped calibrate the first-stage data (as described in Section \autoref{subsec:post}).

\begin{figure}[H]
    \centering
    \includegraphics[width=0.95\linewidth]{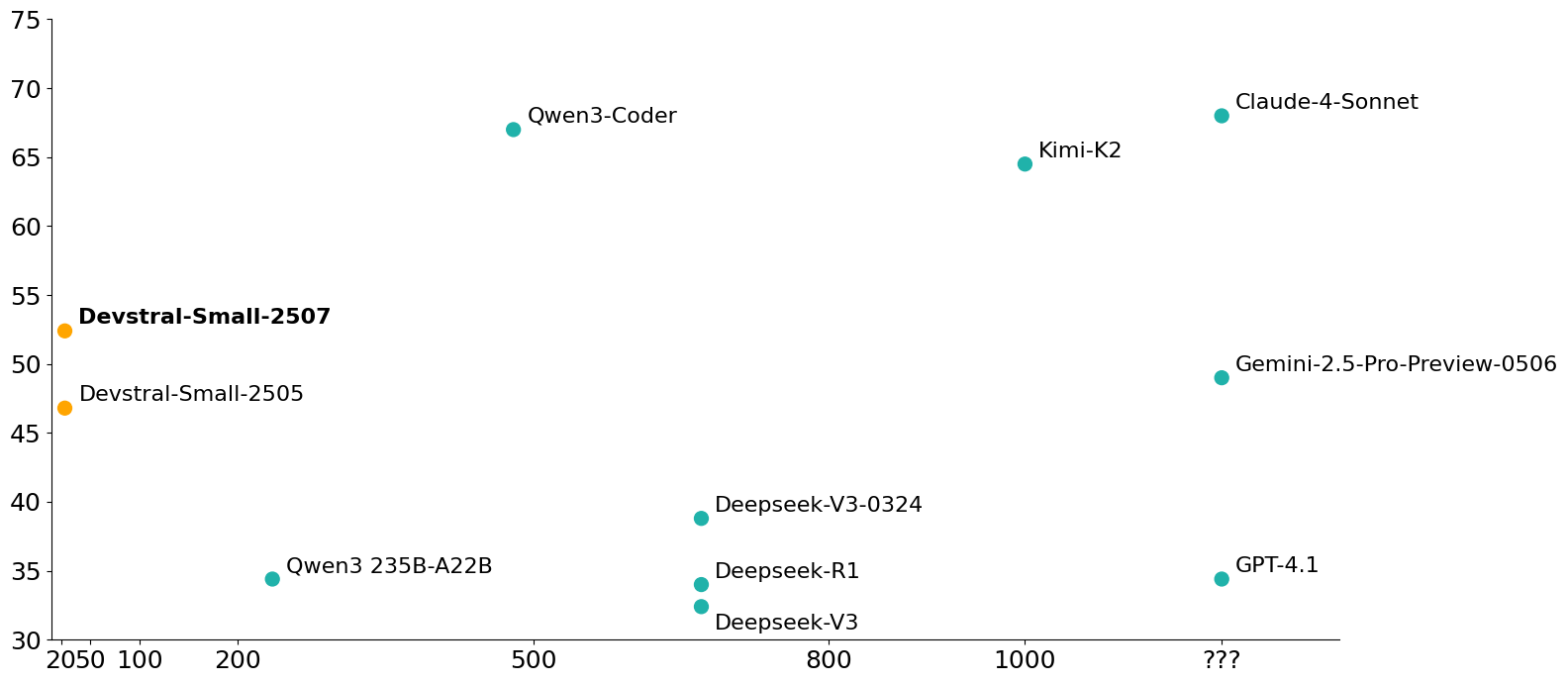}
    \caption{Devstral-2507}
    \label{fig:dev_1_1}
\end{figure}

These improvements resulted in \oursJuly. In Figure \autoref{fig:dev_1_1}, we present the performance versus model size for our models and various other popular models. 
We observe a significant performance improvement between the first and second versions of \oursSmall. These positive results highlight a clear path forward for developing better models, where high-quality data play a crucial role, as seen in many other LLM applications. Additionally, we include two large open-source Mixture of Experts (MoE) models: Kimi K2, with 1 trillion parameters, and Qwen3-Coder, with 480 billion parameters, both released after \oursJuly.

\section{Conclusion}
In this technical report, we introduce \ours, a high-performance open-source model for agentic coding. It remains the best of its weight class, being the most capable model below 100B parameters and easily deployable on device.

\ours is a lightweight 24B model, which learned to use scaffolds to inspect, edit, enhance, and fix code segments within code bases. It is fast and deployable on local devices.

\section*{Core Contributors}
Baptiste Rozière, Graham Neubig, Gauthier Guinet,  Guillaume Lample, Joachim Studnia, Kush Jain, Luyu Gao, Siddharth Gandhi, Wen-Ding Li, Xingyao Wang
\section*{Contributors}
Abhinav Rastogi, Adam Yang, Albert Q. Jiang, Alexander H. Liu, Alexandre Sablayrolles, Amélie Héliou, Amélie Martin, Anmol Agarwal,
Andy Ehrenberg, Andy Lo, Antoine Roux, Arthur Darcet, Arthur Mensch, Baptiste Bout, Baptiste Rozière, Baudouin De Monicault,
Chris Bamford, Christian Wallenwein, Christophe Renaudin, Clémence Lanfranchi, Clément Denoix, Corentin Barreau,
Darius Dabert Devon Mizelle, Diego de las Casas, Elliot Chane-Sane, Emilien Fugier, Emma Bou Hanna,
Gabrielle Berrada, Gauthier Delerce, Georgii Novikov, Guillaume Martin, Himanshu Jaju,
Jan Ludziejewski, Jason Rute, Jean-Malo Delignon, JeanHadrien Chabran, Joep Barmentlo, Jonas Amar, Josselin Somerville Roberts, Julien Denize,
Karan Saxena, Karmesh Yadav, Kartik Khandelwal, Khyathi Raghavi Chandu,
Lélio Renard Lavaud, Léonard Blier, Lingxiao Zhao, Louis Martin, Lucile Saulnier,
Marie Pellat, Mathilde Guillaumin, Mathis Felardos, Matthieu Dinot, Maxime Darrin, Maximilian Augustin, Mickaël Seznec,
Neha Gupta, Nikhil Raghuraman, Olivier Duchenne,
Patricia Wang, Patrick von Platen, Patryk Saffer, Paul Jacob, Paul Wambergue, Paula Kurylowicz, Philomène Chagniot, Pierre Stock, Pravesh Agrawal,
Rémi Delacourt, Roman Soletskyi, Romain Sauvestre,
Sagar Vaze, Sanchit Gandhi, Sandeep Subramanian, Shashwat Dalal, Soham Ghosh, Srijan Mishra, Sumukh Aithal, Szymon Antoniak,
Teven Le Scao, Thibaut Lavril, Thibault Schueller, Thomas Foubert, Thomas Robert, Thomas Wang, Timothée Lacroix, Tom Bewley,
Valeriia Nemychnikova, Victor Paltz, Virgile Richard, William Marshall,
Xuanyu Zhang, Yihan Wan, Yunhao Tang
\newpage

\bibliographystyle{plainnat}
\bibliography{references}

\end{document}